\documentstyle[aps,12pt,tighten]{revtex}

\begin{document}
\author{Sheng Li\thanks{%
Corresponding author. Email: lisheng@itp.ac.cn}}
\address{CCAST(World Lab), P.O.Box 8730, Beijing 100080, P. R. China}
\address{Institute of Theoretical Physics, Academia Sinica, P.O. Box 2735, 
Beijing 100080, P. R. China\thanks{Mailing address}}
\author{Yishi Duan\thanks{%
Email: ysduan@lzu.edu.cn}}
\address{Institute of Theoretical Physics, Lanzhou University, Lanzhou 730000, P. R.
China}
\title{$SO(4)$ Monopole As A New Topological Invariant And Its Topological Structure}
\date{November 9, 1998}
\draft
\maketitle

\begin{abstract}
By making use of the decomposition theory of gauge potential, the inner
structure of $SU(2)$ and $SO(4)$ gauge theory is discussed in detail. We
find the $SO(4)$ monopole can be given via projecting the $SO(4)$ gauge
field onto an antisymmetric tensor. This projection fix the coset $%
SU(2)/U(1)\bigotimes SU(2)/U(1)$ of $SO(4)$ gauge group. The generalized
Hopf map is given via a Dirac spinor. Further we prove that this monopole
can be consider as a new topological invariant. Which is composed of two
monopole structures. Local topological structure of the $SO(4)$ monopole is
discussed in detail, which is quantized by winding number. The Hopf indices
and Brouwer degree labels the local property of the monopoles.
\end{abstract}


\section{Introduction}

It is generally believed, as conjectured by Mandelstam\cite{Mandelstam} and
't Hooft, that magnetic monopoles are essential for confinement in QCD and
related gauge theories. For example, compact QED with a lattice cut off is
known to be exactly dual to a Coulomb gas of monopoles, which upon
condensation causes confinement via a dual Meisner effect\cite{Peskin}.
Similarly there is evidence for the role of monopole condensation in the 3-d
Yang-Mills-Higgs (or Georgi-Glashow) model\cite{Polyakov} and in 4-d N=2
supersymmetric Yang-Mills theory\cite{Seiberg}.

Magnetic monopoles arise when the Higgs configuration has non-trivial
topology at spatial infinity. For a theory with gauge group $G$ broken into
residue group $H$, topologically non-trivial configurations are possible
when the second homotopy group of vacuum manifold, namely $\Pi _2(G/H)$, is
non-trivial. Lots of works are done to understand the structure and the
metric properties of monopole solitons. In order to search for a so-called
non-propagating gauge condition in the non-Abelian theory, 't Hooft\cite
{Hooft} suggest that one should use some tensor $X$, that transforms
covariantly under a gauge transformation $\Omega $:
\begin{equation}
X^{\prime }=\Omega X\Omega ^{-1}.
\end{equation}
The eigenvalues of $X$ are gauge invariant. One can search a gauge in which $%
X$ is diagonal
\begin{equation}
X=\left(
\begin{array}{lll}
\lambda _1 &  & 0 \\
& \ddots  &  \\
0 &  & \lambda _N
\end{array}
\right) .  \label{eigen}
\end{equation}
This gauge condition produces singularities. The singularities occur when
two or more eigenvalues coincide. The nature of these singularities depend
on the gauge group. When the gauge group is $SU(N)$ in the gauge (\ref{eigen}%
) there are singularities, namely if two eigenvalues $\lambda $ of $X$
coincide, which form isolated points in 3-space. However, if the gauge group
is an orthogonal group $SO(N)$, then at non-singular points (\ref{eigen})
fixes the gauge completely, and the singular points, with two coinciding
eigenvalues, form strings in 3-space. These represent Nielson-Olesen
vortices \cite{Nielsen}. So the point like singularities in 3-space or
monopole of the group $SO(N)$ can not be gotten under this gauge. Therefore
finding gauge condition with only point-like singularities for $SO(N)$ is
crucial.

In this paper, we find the gauge condition for $SO(4)$ monopole can be
gotten via projecting the $SO(4)$ gauge field onto an antisymmetric tensor
in close analogy to isospin breaking in the 't Hooft-Polyakov monopole
construction\cite{Hooft2,Polyakov2}. This projection fix the coset $%
SU(2)/U(1)\bigotimes SU(2)/U(1)$ of $SO(4)$ gauge group. The $SO(4)$
monopole obtained is found composed of left and right monopoles, each of
which has the similar structure as that of $SU(2)$ monopole. In the
discussion, the decomposition theory of gauge potential plays an important
role. This theory has been effectively used to study the magnetic monopole
problem in $SU(2)$ gauge theory\cite{Duan1}, the topological gauge theory of
dislocation and disclination in condensed matter physics\cite{Duan5}, the
topological structure of space-time defect\cite{Duan} and the
Gauss-Bonnet-Chern($GBC$) theorem\cite{Duan6,Duan-lisheng}. The inner
structure of $SU(2)$ and $SO(4)$ gauge field in terms of vectors or tensors
gives a deep understanding of construction of the monopoles. By making use
of a Dirac spinor, the $SO(4)$ monopole is proved to be $U(1)\times U(1)$
invariant via a generalized Hopf map $S^3\times S^3\rightarrow S^2\times S^2$%
. Further we show that the $SO(4)$ monopole can be considered as a new
topological invariant.{\em \ }Using the so-called $\phi $-mapping method, it
is verified that the monopole density take the form of a generalized
function $\delta \left( {\bf \phi }\right) $. The positions of the monopoles
are determined by the zeroes of the self-dual or anti-self-dual part of an
antisymmetric tensor. The monopole density is further found as
multi-monopoles system by detailed the structure of delta function, which is
labeled by the Hopf index the Brouwer degree of the antisymmetric tensor
field. Moreover, the direct connection between monopole charges and Winding
number is obtained. In this construction the $SO(4)$ monopole charge density
current is proved taken the same form as the current density of a system of
multi classical point particles.

The organization of the present paper is as follows: In section 2 we start
from the decomposition theory of $SU(2)$ gauge potential and the
construction of $SU(2)$ magnetic monopole for the convenience of the later
discussion of $SO(4)$ monopole. In section 3, the decomposition theory of $%
SO(4)$ gauge potential is given. Using the inner structure of $SO(4)$ gauge
field, we give the construction of the $SO(4)$ monopole. Then in section 4,
we show the $SO(4)$ monopole can be consider as a new topological invariant.
A detailed discussion of the topological structure of the monopole is given
in Section 5. At last, we give a short summary and some concluding remarks
in section 6.

\section{Decomposition theory of $SU(2)$ gauge potential and magnetic
monopole}

Now we firstly give the decomposition theory of $SU(2)$ gauge potential for
the convenience of further discussion of the $SO(4)$ monopoles. The
decomposition theory of gauge potential is a powerful tool in the research
of the topology of gauge field theory. One of the essential features of this
theory is the gauge potential and gauge field possess inner structure, or in
other words, they can be composed of some element fields. For the vector
fields and tensor fields carry important information of topology, the inner
structure of the gauge field in terms of vector fields or tensor fields can
reveal the topological properties of the gauge theory more deeply and
directly. In this section, it will be seen that the $SU(2)$ magnetic
monopole can be induced from decomposition theory naturally.

Let $n$ be an unit $SU(2)$ Lie algebra vector

\begin{equation}
n=n^AI_A\ ,\qquad A=1,2,3\ ;  \label{su2-n}
\end{equation}
and
\begin{equation}
n^An^A=1,  \label{2}
\end{equation}
in which $I_A$ is the generator of the group $SU(2)$, which satisfies the
commutation relation
\begin{equation}
\lbrack I_A,I_B]=-\epsilon ^{ABC}I_C.  \label{su2-communication}
\end{equation}
The covariant derivative 1-form of $n$ is given by
\begin{equation}
D_{SU(2)}n=dn-[\omega _{SU(2)},n],  \label{su2-Dn}
\end{equation}
where $\omega _{SU(2)}$ is the $SU(2)$ gauge potential 1-form:
\begin{equation}
\omega _{SU(2)}=\omega _{SU(2)}^AI_A\ ,  \label{su2-w}
\end{equation}
and
\begin{equation}
\omega _{SU(2)}^A=\omega _{SU(2)\mu }^Adx^\mu \qquad \mu =0,1,2,3.  \label{5}
\end{equation}
The gauge field $2$-form is
\begin{equation}
F_{SU(2)}=F_{SU(2)}^AI_A=d\omega _{SU(2)}-\omega _{SU(2)}\wedge \omega
_{SU(2)},  \label{su2-F}
\end{equation}
and
\begin{eqnarray}
F_{SU(2)}^A &=&\frac 12F_{SU(2)\mu \nu }^Adx^\mu \wedge dx^v  \nonumber \\
&=&d\omega _{SU(2)}^A+\frac 12\epsilon ^{ABC}\omega _{SU(2)}^B\wedge \omega
_{SU(2)}^C.
\end{eqnarray}
Let $m=m^AI_A,\,l=l^AI_A$ be another two unit $SU(2)$ vectors orthonormal to
$n$ and satisfy
\begin{equation}
n^A=\epsilon ^{ABC}m^Bl^C,
\end{equation}
i.e. $n,\,m,\,l$ are orthonormal to each other. It can be proved that the $%
SU(2)$ gauge potential can be decomposed by $n,\,m,\,l$ as

\begin{equation}
\omega _{SU(2)}^A=\epsilon ^{ABC}(dn^Bn^C-D_{SU(2)}n^Bn^C)-n^AA,
\end{equation}
where
\begin{equation}
A=dm^Al^A-D_{SU(2)}m^Al^A.
\end{equation}

From simple caculus, the curvature (gauge field) $2$-form $F_{SU(2)}^A$
becomes
\begin{eqnarray}
F_{SU(2)}^A &=&n^AdA+dn^A\wedge A+\epsilon ^{ABC}n^BdD_{SU(2)}n^C  \nonumber
\\
&&-\frac 12\epsilon ^{ABC}dn^B\wedge dn^C+\frac 12\epsilon
^{ABC}D_{SU(2)}n^B\wedge D_{SU(2)}n^C.  \label{su2gaugefield}
\end{eqnarray}
For $n^A$, $m^A$, $l^A$ form a basis of the vector space. We have
\begin{equation}
F_{SU(2)}^A=(F_{SU(2)}^Bn^B)n^A+(F_{SU(2)}^Bm^B)m^A+(F_{SU(2)}^Bl^B)l^A.
\label{su2gaugefield-all}
\end{equation}
Subsitituting (\ref{su2gaugefield}) into (\ref{su2gaugefield-all}), $%
F_{SU(2)}^B$ is expressed as
\begin{eqnarray}
F_{SU(2)}^A &=&(dA(n)-\frac 12\epsilon ^{DBC}n^Ddn^B\wedge dn^C+\frac 12%
\epsilon ^{DBC}n^DD_{SU(2)}n^B\wedge D_{SU(2)}n^C)n^A  \nonumber \\
&&+(dA(m)-\frac 12\epsilon ^{DBC}m^Ddm^B\wedge dm^C+\frac 12\epsilon
^{DBC}m^DD_{SU(2)}m^B\wedge D_{SU(2)}m^C)m^A  \nonumber \\
&&+(dA(l)-\frac 12\epsilon ^{DBC}l^Ddl^B\wedge dl^C+\frac 12\epsilon
^{DBC}l^DD_{SU(2)}l^B\wedge D_{SU(2)}l^C)l^A.  \label{su2gaugefield-3}
\end{eqnarray}
The above decomposition and inner structure of the $SU(2)$ gauge field is
strict and without any gauge condition. It can be proved this inner
structure has global properties\cite{Duan-lisheng}, which is independent of
the choice of coordinate. From the formula (\ref{su2gaugefield-3}), it's
clear that the $SU(2)$ gauge field is composed of three monopoles on three
orthonormal directions. Therefore, arbitrary direction projection of the
gauge field will give a monopole structure.

In general, the gauge field of $SU(2)$ magnetic monopole can be defined as
\begin{eqnarray}
\tilde{F} &=&F_{SU(2)}^An^A-\frac 12\epsilon ^{ABC}n^AD_{SU(2)}n^B\wedge
D_{SU(2)}n^C  \nonumber \\
&=&n^AdA-\frac 12\epsilon ^{ABC}n^Adn^B\wedge dn^C.  \label{su2-monopole}
\end{eqnarray}
This projection can be considered as a gauge condition which fix the gauge
field on a given direction $n^A$. In this form, the maximal abelian
projection is given. Use the first pair of Maxwell's equation
\begin{eqnarray}
\partial _\nu F^{\mu \nu } &=&-4\pi j^\mu ; \\
F^{\mu \nu } &=&\frac 12\varepsilon ^{\mu \nu \lambda \rho }F_{\lambda \rho
}.
\end{eqnarray}
The magnetic charge current is defined as
\begin{equation}
j^\mu =\frac 1{8\pi }\varepsilon ^{\mu \nu \lambda \rho }\varepsilon
^{ABC}\partial _\nu n^A\partial _\lambda n^B\partial _\rho n^C.
\end{equation}
This expression is of a topological nature since it is conserved
automatically
\begin{equation}
\partial _\mu j^\mu =\frac 1{8\pi }\varepsilon ^{\mu \nu \lambda \rho
}\varepsilon ^{ABC}\partial _\mu (\partial _\nu n^A\partial _\lambda
n^B\partial _\rho n^C)\equiv 0
\end{equation}
due to a complete antisymmetricity of the Levi-Civit\`{a} symbol. Hence the
total magnetic charge $Q_m$
\begin{equation}
Q_m=\int j^\mu dV_\mu   \label{charge}
\end{equation}
is a purely topological quantity.

Now we see the decomposition theory of $SU(2)$ gauge potential gives the
construction of magnetic monopole in a nature way. In next section we will
show how to construct the $SO(4)$ monopole using the decomposition theory of
$SO(4)$ gauge potential.

\section{Inner structure of $SO(4)$ gauge field and $SO(4)$ monopole}

Now let us consider the $SO(4)$ gauge theory. Let the $4$-dimensional Dirac
matrix $\gamma _a$ ($a=1,2,3,4$) be the basis of the Clifford algebra which
satisfies
\begin{equation}
\gamma _a\gamma _b+\gamma _b\gamma _a=2\delta _{ab}.
\label{clliford-communication}
\end{equation}
The antisymmetric tensor field $\phi ^{ab}$ on ${\bf M}$ can be expressed in
the following matrix form
\begin{equation}
\phi =\frac 12\phi ^{ab}I_{ab},
\end{equation}
in which $I_{ab}$ is the generator of the group $SO(4)$%
\begin{equation}
I_{ab}=\frac 14[\gamma _a,\gamma _b].  \label{so4-generator}
\end{equation}
Similarly, the spin connection (gauge potential) $1$-form and curvature
(gage field) $2$-form can be expressed as
\begin{equation}
\omega =\frac 12\omega ^{ab}I_{ab},\quad \quad F=\frac 12F^{ab}I_{ab}.
\end{equation}

It is well known that the spin representation of $SO(4)$ group is hormorphic
to the direct product of the representations of two $SU(2)$ group
\begin{equation}
so(4)\cong su_L(2)\otimes su_R(2).
\end{equation}
Therefore we can get $SO(4)$ monopole via fixing the coset $%
SU_L(2)/U_L(1)\otimes SU_R(2)/U_R(1)$ which should be composed of the left
and right monopoles
\begin{equation}
\tilde{F}=\tilde{F}_L+\tilde{F}_R.
\end{equation}
In the following, we will show how to achieve this goal.

The generators of $SO(4)$ group can be divided into two terms. Each term is
a generator of $SU(2)$ group. Define
\begin{eqnarray}
I_L^1 &=&\frac 12I_{23}(1-\gamma _5)\quad \quad I_R^1=\frac 12%
I_{23}(1+\gamma _5); \\
I_L^2 &=&\frac 12I_{31}(1-\gamma _5)\quad \quad I_R^2=\frac 12%
I_{31}(1+\gamma _5); \\
I_R^3 &=&\frac 12I_{12}(1-\gamma _5)\quad \quad I_R^3=\frac 12%
I_{12}(1+\gamma _5),
\end{eqnarray}
in which $\gamma _5=\gamma _1\gamma _2\gamma _3\gamma _4$. It can be proved $%
I_{L(R)}^A(A=1,2,3)$ satisfy the commutation relation of group $SU(2)$%
\begin{equation}
I_{L(R)}^A=-\epsilon ^{ABC}[I_{L(R)}^B,I_{L(R)}^C],
\end{equation}
and
\begin{equation}
\lbrack {I_L^A,I_R^B]}=0.  \label{comutivity}
\end{equation}
Therefore $I_{L(R)}^A$ are the generators of the groups $SU(2)_{L(R)}$, and
they are the basis of $SU(2)_{L(R)}$ Lie algebra spaces.

Arbitrary antisymmetric tensor field $\phi ^{ab}$ can be decomposed as
\begin{equation}
\phi ^{ab}=\phi _L^{ab}+\phi _R^{ab},
\end{equation}
where
\begin{equation}
\phi _L^{ab}=\frac 12(\phi ^{ab}+\frac 12\epsilon ^{abcd}\phi ^{cd})\quad
\quad \phi _R^{ab}=\frac 12(\phi ^{ab}-\frac 12\epsilon ^{abcd}\phi ^{cd})
\end{equation}
are the self-dual and anti-self-dual parts of $\phi ^{ab}$. Define
\begin{equation}
\phi _L^1=\phi ^{23}+\phi ^{14},\quad \phi _L^2=\phi ^{31}+\phi ^{24},\quad
\phi _L^3=\phi ^{12}+\phi ^{34};
\end{equation}
\begin{equation}
\phi _R^1=\phi ^{23}-\phi ^{14},\quad \phi _R^2=\phi ^{31}-\phi ^{24},\quad
\phi _R^3=\phi ^{12}-\phi ^{34}.
\end{equation}
We can rewrite $\phi $ as
\begin{equation}
\phi =\frac 12\phi ^{ab}I_{ab}=\phi _L+\phi _R,
\end{equation}
and
\begin{equation}
||\phi _{L(R)}||=\phi _{L(R)}^A\phi _{L(R)}^A=\phi _{L(R)}^{ab}\phi
_{L(R)}^{ab},
\end{equation}
in which $\phi _L$ and $\phi _R$ are just the components of $\phi $
correspond to $SU_L(2)$ and $SU_R(2)$ Lie algebra
\begin{eqnarray}
\phi _L &=&\frac 12\phi (1-\gamma _5)=\phi _L^AI_L^A, \\
\phi _R &=&\frac 12\phi (1+\gamma _5)=\phi _R^AI_R^A.
\end{eqnarray}
With the similar decomposition as above it can be verified that
\begin{equation}
D\phi =D\phi _L+D\phi _R=d\phi _L-[\omega _L,\phi _L]+d\phi _R-[\omega
_R,\phi _R].
\end{equation}
For the independence of $I_L$ and $I_R$ we have
\begin{equation}
D\phi _{L(R)}=d\phi _{L(R)}-[\omega _{L(R)},\phi _{L(R)}],
\label{lrderevitive}
\end{equation}
or in components representation
\begin{equation}
D_{L(R)}\phi _{L(R)}^A=d\phi _{L(R)}^A-\epsilon ^{ABC}\omega _{L(R)}^B\phi
_{L(R)}^C.
\end{equation}
in which
\begin{equation}
\omega _L=\omega _L^AI_L^A\quad \quad \quad \omega _R=\omega _R^AI_R^A
\end{equation}
is the gauge potential of $SU(2)_{L(R)}$ gauge field.

Similarly the curvature $2$-form can be decomposed as
\begin{equation}
F=F_L+F_R,
\end{equation}
and
\begin{equation}
F_L=d\omega _L-\omega _L\wedge \omega _L\quad \quad F_R=d\omega _R-\omega
_R\wedge \omega _R.
\end{equation}
$F_{L(R)}=F_{L(R)}^AI_{L(R)}^A$ are the curvature $2$-forms of $SU(2)_{L(R)}$
gauge field.

Define an antisymmetric tensor
\begin{equation}
n^{ab}=\frac{\phi _L^{ab}}{||\phi _L||}+\frac{\phi _R^{ab}}{||\phi _R||},
\label{tensor-unity}
\end{equation}
i.e.
\begin{equation}
n_L^{ab}=\frac{\phi _L^{ab}}{||\phi _L||},\quad \quad \quad \quad \quad
n_R^{ab}=\frac{\phi _R^{ab}}{||\phi _R||}.
\end{equation}
which has the properties
\begin{equation}
n^{ab}n^{ab}=2\quad \quad \varepsilon ^{abcd}n^{ab}n^{cd}=0.
\end{equation}
Then
\begin{equation}
n=\frac 12n^{ab}I_{ab}=n_L+n_R
\end{equation}
and
\begin{equation}
n_L=\frac 1{2||\phi _L||}\phi (1-\gamma _5)=n_L^AI_L^A,  \label{n-d}
\end{equation}
\begin{equation}
n_R=\frac 1{2||\phi _R||}\phi (1+\gamma _5)=n_R^AI_R^A.
\end{equation}
It naturally guarantees the constraint
\begin{equation}
n_{L(R)}^An_{L(R)}^A=1.  \label{l-r-unity}
\end{equation}
i.e. $n_{L(R)}^A$ are the $SU(2)_{L(R)}$ unit vectors.

From the decomposition formula of $SU(2)$ gauge potential (\ref
{su2gaugefield}) we get
\begin{eqnarray}
F_{L(R)}^A &=&n_{L(R)}^AdA_{L(R)}+dn_{L(R)}^A\wedge A+\epsilon
^{ABC}n_{L(R)}^BdD_{L(R)}n_{L(R)}^C  \nonumber \\
&&-\frac 12\epsilon ^{ABC}dn_{L(R)}^B\wedge dn_{L(R)}^C+\frac 12\epsilon
^{ABC}D_{L(R)}n_{L(R)}^B\wedge D_{L(R)}n_{L(R)}^C.  \label{so4-Fd}
\end{eqnarray}
Now we can define the left and right $SO(4)$ monopoles as
\begin{eqnarray}
\tilde{F}_{L(R)} &=&F_{L(R)}^An_{L(R)}^A-\frac 12\epsilon
^{ABC}n_{L(R)}^AD_{L(R)}n_{L(R)}^B\wedge D_{L(R)}n_{L(R)}^C  \nonumber \\
&=&n_{L(R)}^AdA_{L(R)}-\frac 12\epsilon ^{ABC}n_{L(R)}^Adn_{L(R)}^B\wedge
dn_{L(R)}^C.
\end{eqnarray}
Through simple computation, we can get the gauge field of $SO(4)$ monopole
\begin{eqnarray}
\tilde{F} &=&\tilde{F}_L+\tilde{F}_R  \nonumber \\
&=&F^{ab}n^{ab}+n^{ab}Dn^{ac}\wedge Dn^{cb}  \nonumber \\
&=&n^{ab}dn^{ac}\wedge dn^{cb}+d(A_L+A_R).  \label{so4monopole}
\end{eqnarray}
In this way the coset $SU(2)_L/U(1)_L\bigotimes SU(2)_R/U(1)_R$ of $SO(4)$
group is fixed. The corresponding monopole charge current is
\begin{eqnarray}
j^\mu  &=&-\frac 1{8\pi }\varepsilon ^{\mu \nu \lambda \rho }\partial _\nu
\tilde{F}_{\lambda \rho }  \nonumber \\
&=&-\frac 1{4\pi }\varepsilon ^{\mu \nu \lambda \rho }\partial _\nu
n^{ab}\partial _\lambda n^{ac}\partial _\rho n^{cb},
\end{eqnarray}
which is also conserved automatically
\begin{equation}
\partial _\mu j^\mu =-\frac 1{4\pi }\varepsilon ^{\mu \nu \lambda \rho
}\partial _\mu (\partial _\nu n^{ab}\partial _\lambda n^{ac}\partial _\rho
n^{cb})\equiv 0
\end{equation}
The total monopole charge
\begin{equation}
Q_m=\int j^\mu dV_\mu =-\frac 1{4\pi }\int dn^{ab}\wedge dn^{ac}\wedge
dn^{cb}
\end{equation}
is a purely topological quantity.

On another hand, the inner structure of $SO(4)$ gauge field can be gotten
from the decomposition theory of gauge potential as
\begin{eqnarray}
F^{ab} &=&dn^{ac}\wedge dn^{cb}-Dn^{ac}\wedge
Dn^{cb}-n^{ac}dDn^{cb}+n^{bc}dDn^{ca}  \nonumber \\
&&+dA_Ln_L^{ab}+dA_Rn_R^{ab}-A_L\wedge Dn_L^{ab}-A_R\wedge Dn_R^{ab}.
\label{so4gaugefield-tensor}
\end{eqnarray}
Then use three antisymmetric orthogornal antisymmetric tensors $n^{ab}$, $%
m^{ab}$, $l^{ab}$%
\begin{eqnarray}
n^{ab}m^{ab} &=&n^{ab}l^{ab}=m^{ab}l^{ab}=0,  \nonumber \\
\varepsilon ^{abcd}n^{ab}m^{cd} &=&\varepsilon
^{abcd}n^{ab}l^{cd}=\varepsilon ^{abcd}m^{ab}l^{cd}=0
\end{eqnarray}
and
\begin{eqnarray}
n^{ab}n^{ab} &=&m^{ab}m^{ab}=l^{ab}l^{ab}=2,  \nonumber \\
\varepsilon ^{abcd}n^{ab}n^{cd} &=&\varepsilon
^{abcd}m^{ab}m^{cd}=\varepsilon ^{abcd}l^{ab}l^{cd}=0,
\end{eqnarray}
we have
\begin{eqnarray}
F^{ab} &=&\frac 12(F^{cd}n^{cd})n^{ab}+\frac 12(F^{cd}m^{cd})m^{ab}+\frac 12%
(F^{cd}l^{cd})l^{ab}  \nonumber \\
&+&\frac 12(F^{cd}n^{*cd})n^{*ab}+\frac 12(F^{cd}m^{*cd})m^{*ab}+\frac 12%
(F^{cd}l^{*cd})l^{*ab},  \label{gaugefield-3}
\end{eqnarray}
in which $n^{*ab}$, $m^{*ab}$and $l^{*ab}$ are the dual tensors of $n^{ab}$,
$m^{ab}$, $l^{ab}$%
\begin{equation}
n^{*ab}=\frac 12\varepsilon ^{abcd}n^{cd}\quad m^{*ab}=\frac 12\varepsilon
^{abcd}m^{cd}\quad l^{*ab}=\frac 12\varepsilon ^{abcd}l^{cd}.
\end{equation}
Substituting the inner structure (\ref{so4gaugefield-tensor}) into (\ref
{gaugefield-3}), we get
\begin{eqnarray}
F^{ab} &=&\frac 12(n^{dc}dn^{de}\wedge dn^{ec}-n^{dc}Dn^{de}\wedge
Dn^{ec}+dA_L(n)+dA_R(n))n^{ab}  \nonumber \\
&&+\frac 12(m^{dc}dm^{de}\wedge dm^{ec}-m^{dc}Dm^{de}\wedge
Dm^{ec}+dA_L(m)+dA_R(m))m^{ab}  \nonumber \\
&&+\frac 12(l^{dc}dl^{de}\wedge dl^{ec}-l^{dc}Dl^{de}\wedge
Dl^{ec}+dA_L(l)+dA_R(l))l^{ab}  \nonumber \\
&&+\frac 12(n^{*dc}dn^{*de}\wedge dn^{*ec}-n^{*dc}Dn^{*de}\wedge
Dn^{*ec}+dA_L(n^{*})+dA_R(n^{*}))n^{*ab}  \nonumber \\
&&+\frac 12(m^{*dc}dm^{*de}\wedge dm^{*ec}-m^{*dc}Dm^{*de}\wedge
Dm^{*ec}+dA_L(m^{*})+dA_R(m^{*}))m^{*ab}  \nonumber \\
&&+\frac 12(l^{*dc}dl^{*de}\wedge dl^{*ec}-l^{*dc}Dl^{*de}\wedge
Dl^{*ec}+dA_{L^{*}}(l^{*})+dA_R(l^{*}))l^{*ab}.
\end{eqnarray}
So we see from the above formula the gauge field is composed of six $SO(4)$
monopoles in three orthogornal tensor directions and their dual directions.
The monopole projection onto the dual tensor of $n^{ab}$ is just the
difference of the left and right monopole projection onto $n^{ab}$%
\begin{equation}
\tilde{F}(n^{*})=\tilde{F}(n)_L-\tilde{F}(n)_R.
\end{equation}

Given a Dirac spinor
\begin{equation}
\psi =\left(
\begin{array}{c}
\psi _L \\
\psi _R
\end{array}
\right) .
\end{equation}
Satisfying
\begin{equation}
\psi _L^{\dagger }\psi _L=1\quad \quad \psi _R^{\dagger }\psi _R=1.
\label{spinor-unity}
\end{equation}
Define the antisymmetric tensor $n^{ab}$ by
\begin{equation}
n^{ab}=\psi ^{\dagger }I_{ab}\psi .  \label{tensor-spinor}
\end{equation}
It is easy to prove $n^{ab}$ can be regard as the antisymmetric tensor
defined in (\ref{tensor-unity}) satisfying
\begin{equation}
n^{ab}n^{ab}=2\quad \quad \varepsilon ^{abcd}n^{ab}n^{cd}=0.
\end{equation}
The self-dual part of $n^{ab}$ is
\begin{equation}
n_L^{ab}=\frac 12(\psi ^{\dagger }I_{ab}\psi +\frac 12\varepsilon
^{abcd}\psi ^{\dagger }I_{cd}\psi ).
\end{equation}
It's easy to prove the above equation can be rewritten as
\begin{eqnarray}
n_L^{ab} &=&\frac 12\psi ^{\dagger }I_{ab}(1-\gamma ^5)\psi   \nonumber \\
&=&\frac 12\left(
\begin{array}{c}
\psi _L \\
0
\end{array}
\right) ^{\dagger }I_{ab}(1-\gamma ^5)\left(
\begin{array}{c}
\psi _L \\
0
\end{array}
\right) .
\end{eqnarray}
Similarly we have
\begin{equation}
n_R^{ab}=\frac 12\left(
\begin{array}{c}
0 \\
\psi _R
\end{array}
\right) ^{\dagger }I_{ab}(1+\gamma ^5)\left(
\begin{array}{c}
0 \\
\psi _R
\end{array}
\right) .
\end{equation}
The $SO(4)$ monopole gauge field $\tilde{F}$ (\ref{so4monopole}) becomes
\begin{equation}
\tilde{F}=id\psi ^{\dagger }d\psi +d(A_L+A_R)
=id\psi _L^{\dagger }d\psi _L+id\psi
_R^{\dagger }d\psi _R+d(A_L+A_R).
\end{equation}
Therefore, given a transformation
\begin{equation}
\psi ^{\prime }=\left(
\begin{array}{cc}
e^{i\theta }I & 0 \\
0 & e^{i\varphi }I
\end{array}
\right) \psi .
\end{equation}
It is easy to prove the gauge field 2-form $\tilde{F}$ is invariant under
this transformation, therefore $\tilde{F}$ given here is $U(1)\times U(1)$
invariant.

The condition (\ref{spinor-unity}) give a direct product of two $S^3$ sphere
\begin{equation}
S^3\times S^3,
\end{equation}
and the condition (\ref{l-r-unity}) gives a product of two $S^2$ sphere
\begin{equation}
S^2\times S^2.
\end{equation}
The definition (\ref{tensor-spinor}) actually gives a generalized Hopf map%
\cite{Ryder,Minami}
\begin{equation}
S^3\times S^3\rightarrow S^2\times S^2.
\end{equation}

\section{$SO(4)$ monopole as a new topological invariant}

A topological invariant should has two properties. First, it has to be gauge
invariant. And second, it should be gauge independent, or in other words, it
has the same value under arbitrary gauge conditions. In the follows we will
prove the gauge field of $SO(4)$ monopole indeed possess these two
properties.

Firstly, under arbitrary gauge transformation, the curvature 2-form $F^{ab}$
transform covariantly as
\begin{equation}
F^{\prime ab}=S^{ac}F^{cd}S^{db}
\end{equation}
and the covariant derivative 1-form of the antisymmetric tensor $n^{ab}$
transform covariantly also
\begin{equation}
D^{\prime }n^{\prime ab}=S^{ac}Dn^{cd}S^{cb},
\end{equation}
where $S^{ab}$ is the element of the $SO(4)$ gauge group satisfying
\begin{equation}
S^{ac}S^{cb}=\delta ^{ab}.
\end{equation}
Hence we can prove the new topological invariant $\tilde{F}$ is $SO(4)$
gauge invariant, i.e.
\begin{equation}
\tilde{F}^{\prime }=\tilde{F}.
\end{equation}

Then let us prove the gauge independence property of $\tilde{F}$. Given two $%
SO(4)$ spin connection $\omega _0$ and $\omega _1$, the corresponding
monopole gauge field 2-forms are
\begin{equation}
\tilde{F}(\omega _0)=n^{ab}dn^{ac}\wedge dn^{cb}+dA_L(\omega _0)+dA_R(\omega
_0)
\end{equation}
and
\begin{equation}
\tilde{F}(\omega _1)=n^{ab}dn^{ac}\wedge dn^{cb}+dA_L(\omega _1)+dA_R(\omega
_1).
\end{equation}
Then the difference between $\tilde{F}(\omega _0)$ and $\tilde{F}(\omega _1)$
is just a exact 2-form
\begin{equation}
\tilde{F}(\omega _1)-\tilde{F}(\omega _0)=d(A_L(\omega _1)+A_R(\omega
_1)-A_L(\omega _0)-A_R(\omega _0)).
\end{equation}
Thus the integrals of $\tilde{F}(\omega _0)$ and $\tilde{F}(\omega _1)$ over
a closed 2-dimensional surface give the same results and hence we prove $%
\tilde{F}$ is independent of the gauge potential. The above two properties,
i.e. gauge invariance and independence of gauge potential, make the integral
of our monopole gauge field $2$-form over a closed 2-dimensional surface to
a topological invariant.

At last we get the new topological invariant $Q\,$%
\begin{equation}
Q=\frac 1{4\pi }\int_{\Sigma ^2}\tilde{F}=\frac 1{4\pi }\int_{V^3}d\tilde{F}=%
\frac 1{4\pi }\int_{V^3}dn^{ab}\wedge dn^{ac}\wedge dn^{cb},
\end{equation}
in which $\Sigma ^2=\partial V^3$.

In another hand the monopole charge is given by integral
\begin{eqnarray}
Q_m &=&\int_Vj^\mu dV_\mu   \nonumber \\
&=&-\frac 1{8\pi }\int_V\varepsilon ^{\mu \nu \lambda \rho }\partial _\nu
\tilde{F}_{\lambda \rho }dV_\mu   \nonumber \\
&=&-\frac 1{4\pi }\int_Vd\tilde{F}.
\end{eqnarray}
Now we see the monopole charge and the new topological invariant is the same
in fact but a minus sign
\begin{equation}
Q_m=-Q.
\end{equation}

\section{The local topological structure of the $SO(4)$ monopole}

Define left and right monpole charges as
\begin{equation}
Q_{mL(R)}=\int_Vj_{L(R)}^\mu dV_\mu =\frac 1{8\pi }\int_V\epsilon
^{ABC}dn_{L(R)}^A\wedge dn_{L(R)}^B\wedge dn_{L(R)}^C.  \label{W-d}
\end{equation}
The total monpole charge is
\begin{equation}
Q_m=Q_{mL}+Q_{mR}.
\end{equation}

Let $y=(u^1,u^2,u^3,\tau )$ be another term of the coordinate of $M$ and $%
u=(u^1,u^2,u^3)$ be the intrinsic coordinate of $V$. For the coordinate
component $v$ does not belong to $V$. Then
\begin{eqnarray}
\Omega _{mL(R)} &=&\frac 1{8\pi }\int_V\epsilon ^{ABC}\partial
_in_{L(R)}^A\partial _jn_{L(R)}^B\partial _kn_{L(R)}^Cdu^i\wedge du^j\wedge
du^k  \nonumber \\
&=&\frac 1{8\pi }\int_V\epsilon ^{ijk}\epsilon ^{ABC}\partial
_in_{L(R)}^A\partial _jn_{L(R)}^B\partial _kn_{L(R)}^Cd^3u,
\end{eqnarray}
where $i,j,k=1,2,3$ and $\partial _i=\partial /\partial u^i$. Then the
monpole charge densities can be defined as
\begin{equation}
\rho _{L(R)}=\frac 1{8\pi }\epsilon ^{ijk}\epsilon ^{ABC}\partial
_in_{L(R)}^A\partial _jn_{L(R)}^B\partial _kn_{L(R)}^C.  \label{solid-angle}
\end{equation}
We get
\begin{equation}
Q_{mL(R)}=\int_V\rho _{L(R)}d^3u.
\end{equation}

For the equation (\ref{n-d}), the unit vectors $n_{L(R)}^A(x)$ can expressed
as follows:
\begin{equation}
n_{L(R)}^A=\frac{\phi _{L(R)}^A}{||\phi _{L(R)}||}.
\end{equation}
Hence
\begin{equation}
dn_{L(R)}^A=\frac 1{||\phi _{L(R)}||}d\phi _{L(R)}^A+\phi _{L(R)}^Ad(\frac 1{%
||\phi _{L(R)}||}),
\end{equation}
and
\begin{equation}
\frac \partial {\partial \phi _{L(R)}^A}(\frac 1{||\phi _{L(R)}||})=-\frac{%
\phi _{L(R)}^A}{||\phi _{L(R)}||^3}.
\end{equation}
Substituting above equations into the monopole charge density (\ref
{solid-angle}) we obtain
\begin{eqnarray}
\rho  &=&\frac 1{8\pi }\epsilon ^{ABC}\epsilon ^{ijk}\partial
_i(n_{L(R)}^A\partial _jn_{L(R)}^B\partial _kn_{L(R)}^C)  \nonumber \\
&=&\frac 1{8\pi }\epsilon ^{ABC}\epsilon ^{ijk}\partial _i\frac{\phi
_{L(R)}^A}{||\phi _{L(R)}||^3}\partial _j\phi _{L(R)}^B\partial _k\phi
_{L(R)}^C  \nonumber \\
&=&-\frac 1{8\pi }\epsilon ^{ABC}\epsilon ^{ijk}\frac \partial {\partial
\phi _{L(R)}^D}\frac \partial {\partial \phi _{L(R)}^A}(\frac 1{||\phi
_{L(R)}||})\partial _i\phi _{L(R)}^D\partial _j\phi _{L(R)}^B\partial _k\phi
_{L(R)}^C.
\end{eqnarray}
Define the Jacobian $D(\frac{\phi _{L(R)}}u)$ as
\begin{equation}
\epsilon ^{ABC}D(\frac{\phi _{L(R)}}u)=\epsilon ^{ijk}\partial _i\phi
_{L(R)}^A\partial _j\phi _{L(R)}^B\partial _k\phi _{L(R)}^C.
\end{equation}
By making use of the Laplacian relation in $\phi $-space
\begin{equation}
\partial _A\partial _A\frac 1{||\phi _{L(R)}||}=-4\pi \delta ^3(\phi
_{L(R)}),\quad \partial _A=\frac \partial {\partial \phi _{L(R)}^A},
\end{equation}
we can write the monopole charge density as the $\delta $-like expression
\begin{equation}
\rho _{L(R)}=\delta ^3(\phi _{L(R)})D(\frac{\phi _{L(R)}}u)
\label{density-1}
\end{equation}
and
\begin{equation}
Q_{mL(R)}=\int_V\delta ^3(\phi _{L(R)})D(\frac{\phi _{L(R)}}u)d^3u.
\label{m-charge}
\end{equation}
It obvious that $\rho _{L(R)}$ are non-zero only when $\phi _{L(R)}=0$.

Suppose that $\phi _{L(R)}^A(x)$ ($A=1,2,3)$ possess $K_{L(R)}$ isolated
zeros, according to the implicit function theorem, the solutions of $\phi
_{L(R)}(u^1,u^2,u^3,\tau )=0$ can be expressed in terms of $u=(u^1,u^2,u^3)$
as
\begin{equation}
u^i=z_{L(R)}^i(\tau ),\quad \quad \quad \quad \quad i=1,2,3
\end{equation}
and
\begin{equation}
\phi _{L(R)}^A(z_l^1(\tau ),z_l^2(\tau ),z_l^3(\tau ),\tau )\equiv 0,
\end{equation}
where the subscript $l=1,2,\cdots ,K_{L(R)}$ represents the $l$th zero of $%
\phi _{L(R)}^A$, i.e.
\begin{equation}
\phi _{L(R)}^A(z_{L(R)l}^i)=0,\quad \quad \quad l=1,2,\cdots ,K_{L(R)}.
\end{equation}
It is easy to get the following formula from the ordinary theory of $\delta $%
-function that
\begin{equation}
\delta ^3(\phi _{L(R)})D(\frac{\phi _{L(R)}}u)=\sum_{i=1}^{K_{L(R)}}\beta
_{L(R)l}\eta _{L(R)l}\delta ^3(u-z_{L(R)l})
\end{equation}
in which
\begin{equation}
\eta _{L(R)l}=signD(\frac{\phi _{L(R)}}u)|_{x=z_{L(R)l}}=\pm 1,
\end{equation}
is the Brouwer degree of $\phi $-mapping and $\beta _{L(R)l}$ are positive
integers called the Hopf index of map $\phi _{L(R)}$ which means while the
point $x$ covers the region neighboring the zero $x=z_{L(R)l}$ once, $\phi
_{L(R)}$ covers the corresponding region $\beta _{L(R)l}$ times. Therefore
the slid angle density becomes
\begin{equation}
\rho _{L(R)}=\sum_{l=1}^{K_{L(R)}}\beta _{L(R)l}\eta _{L(R)l}\delta
^3(u-z_{L(R)l})  \label{density-d}
\end{equation}
and
\begin{equation}
Q_{mL(R)}=\sum_{l=1}^{K_{L(R)}}\beta _{L(R)l}\eta _{L(R)l}\int_V\delta
^3(u-z_{L(R)l})d^3u=\sum_{l=1}^{K_{L(R)}}\beta _{L(R)l}\eta _{L(R)l}.
\end{equation}
Therefore the total $SO(4)$ monopole charge is
\begin{equation}
Q_m=\sum_{l=1}^{K_L}\beta _{Ll}\eta _{Ll}+\sum_{l=1}^{K_R}\beta _{Rl}\eta
_{Rl}.
\end{equation}

We find that (\ref{density-d}) is the exact density of a system of $K_L$ and
$K_R$ classical point-like objects with ``charge'' $\beta _{Ll}\eta _{Ll}$
and $\beta _{Rl}\eta _{Rl}$ in space-time, i.e. the topological structure of
$SO(4)$ monopole charge density formally corresponds to a point-like system.
These point objects may be called topological particles which are identified
with the isolated zero points of vector field $\phi _{L(R)}^A(x)$.

Using similar way it can be proved the monopole charge current density can
be expressed as
\begin{equation}
j^\mu =\sum_{l=1}^{K_L}\beta _{Ll}\eta _{Ll}\delta ^3(u-z_{Ll})\frac{D^\mu (%
\frac{\phi _L}x)}{D(\frac{\phi _L}u)}+\sum_{l=1}^{K_R}\beta _{Rl}\eta
_{Rl}\delta ^3(u-z_{Rl})\frac{D^\mu (\frac{\phi _R}x)}{D(\frac{\phi _R}u)},
\end{equation}
where the $D^\mu (\frac{\phi _{L(R)}}x)$ is the Jacobian vector
\begin{equation}
\epsilon ^{ABC}D^\mu (\frac{\phi _{L(R)}}u)=\epsilon ^{\mu \nu \lambda \rho
}\partial _\nu \phi _{L(R)}^A\partial _\lambda \phi _{L(R)}^B\partial _\rho
\phi _{L(R)}^C.
\end{equation}
It can be proved the general velocity of the $i$th zero is
\begin{equation}
V_l^\mu =\left. \frac{dx^\mu }{d\tau }\right| _{x=z_l}=\left. \frac{D^\mu (%
\frac \phi x)}{D(\frac \phi u)}\right| _{x=z_l},
\end{equation}
then
\begin{eqnarray}
j^\mu  &=&\sum_{l=1}^{K_L}\beta _{Ll}\eta _{Ll}\delta ^3(u-z_{Ll})\frac{%
dx^\mu }{d\tau }  \nonumber \\
&&+\sum_{l=1}^{K_R}\beta _{Rl}\eta _{Rl}\delta ^3(u-z_{Rl})\frac{dx^\mu }{%
d\tau }.
\end{eqnarray}
which is in the same form as the classical current density of the system of $%
K_L+K_R$ point particles moving in the space-time.

On another hand, the winding number of the surface $\Sigma $ and the mapping
$n_{L(R)}$ is defined as\cite{Victor}
\begin{equation}
W_{L(R)}=\oint_\Sigma \frac 1{8\pi }\epsilon ^{ABC}\frac{\phi _{L(R)}^A}{%
||\phi _{L(R)}||^3}d\phi _{L(R)}^B\wedge d\phi _{L(R)}^C,  \label{winding}
\end{equation}
which is equal to the number of times $\Sigma $ encloses (or, wraps around)
the point $\phi _{L(R)}=0$. Hence, the monopole charge is quantized by the
winding numbers
\begin{equation}
Q_{mL(R)}=W_{L(R)}.  \label{windingnumber}
\end{equation}
The winding number $W_{L(R)}$ of the surface $\Sigma $ can be interpreted
or, indeed, defined as the degree of the mappings $\phi _{L(R)}$ onto $%
\Sigma $. By (\ref{m-charge}) and (\ref{density-1})we have
\begin{eqnarray}
Q_{mL(R)} &=&\int_V\delta (\phi _{L(R)})\rho _{L(R)}(\frac{\phi _{L(R)}}u%
)d^3u  \nonumber \\
&=&\deg \phi _{L(R)}\int_{\phi _{L(R)}(V)}\delta (\phi _{L(R)})d^3\phi
_{L(R)}  \nonumber \\
&=&\deg \phi _{L(R)},
\end{eqnarray}
where $\deg \phi _{L(R)}$ are the degrees of map $\phi _{L(R)}:V\rightarrow
\phi _{L(R)}(V).$ Compared above equation with (\ref{windingnumber}), it
shows the degrees of map $\phi _{L(R)}:V\rightarrow \phi _{L(R)}(V)$ is just
the winding number $W_{L(R)}$ of surface $\Sigma $ and map $\phi _{L(R)}$,
i.e.
\begin{equation}
\deg \phi _{L(R)}=W_{L(R)}(\Sigma ,\phi _{L(R)}).
\end{equation}
Then the total $SO(4)$ monopole charge is
\begin{equation}
Q_m=\deg \phi _L+\deg \phi _R=W_L+W_R.
\end{equation}
Divide $V$ by
\begin{equation}
V=\sum_{l=1}^{K_{L(R)}}V_{L(R)l}
\end{equation}
so that $V_{L(R)l}$ includes only one zero $z_{L(R)l}$ of $\phi _{L(R)}$,
i.e. $z_{L(R)l}\in V_{L(R)l}$. The winding number of the surface $\Sigma
_{L(R)l}=\partial V_{L(R)l}$ and the mapping $n_{L(R)}$ is defined as
\begin{equation}
W_{L(R)l}=\oint_{\Sigma _{L(R)l}}\frac 1{8\pi }\epsilon ^{ABC}\frac{\phi
_{L(R)}^A}{||\phi _{L(R)}||^3}d\phi _{L(R)}^B\wedge d\phi _{L(R)}^C,
\end{equation}
which is equal to the number of times $\Sigma _{L(R)l}$ encloses (or, wraps
around) the point $\phi _{L(R)}=0$. It is easy to see that
\begin{equation}
W_{L(R)}=\sum_{l=1}^{K_{L(R)}}W_{L(R)l}
\end{equation}
and
\begin{equation}
|W_{L(R)l}|=\beta _{L(R)l}.
\end{equation}
Then
\begin{equation}
Q_m=\sum_{l=1}^{K_L}W_{Ll}+\sum_{l=1}^{K_L}W_{Rl}.
\end{equation}
Also we can write
\begin{equation}
Q_m=(N_L^{+}+N_R^{+})-(N_L^{-}+N_R^{-}),
\end{equation}
in which $N_{L(R)}^{\pm }$are the sums of the Hopf indexes with respect to $%
\eta _{L(R)}=\pm 1$. We see that while $x$ covers $V$ once, $\phi _{L(R)}^A$
must cover $\phi _{L(R)}(V)$ $N_{L(R)}^{+}$ times with $\eta =1$ and $%
N_{L(R)}^{-}$ times with $\eta =-1$.

\section{Conclusion}

In this paper we have provide a gauge condition via projecting the $SO(4)$
gauge field onto an antisymmetric tensor field to fix the coset $%
SU(2)_L/U(1)_L\bigotimes SU(2)_R/U(1)_R$ of $SO(4)$ group. Under this gauge
condition the $SO(4)$ monopole is constructed which is composed of two
monopoles, namely left and right monopoles. Each of these monopoles has the
similar structure as that of the $SU(2)$ magnetic monopoles. We also
discussed the topological properties of this monopole and find it can be
consider as a new topological invariant. Using a Dirac spinor, we give a
generalized Hopf map $S^3\times S^3\rightarrow S^2\times S^2$. The detailed
local topological structure is given. It's the monopole charge is quantized
topologically by winding number. The Hopf indices and Brouwer degree
characterize the monopoles.

The decomposition theory of gauge potential play an important role in the
discussion. The monopole structure can be induced from the inner structure
of the gauge field easily. It is again a proof the success of the
decomposition theory. We assume that the $SO(N)$ ($N>4)$ monopole structure
may be induced by making use of this theory in analogous way.

\acknowledgments
This research work is supported by National Natural Science Foundation of P.
R. China.


\begin{references}
\bibitem{Mandelstam}  S. Mandelstam, Phys. Rep. C23, 245 (1976).

\bibitem{Hooft}  G. 't Hooft, Nucl. Phys. B190 (1981) 455.

\bibitem{Peskin}  M.E. Peskin, Ann. of Phys. 113 122 (1978).

\bibitem{Polyakov}  A.M. Polyakov Phys. Lett. B59, 82 (1975); Nucl. Phys.
B120 429 (1977).

\bibitem{Seiberg}  N. Seiberg and E. Witten, Nucl. Phys. B426 19 (1994).

\bibitem{Nielsen}  H.B. Nielsen and P. Olesen, Nucl. Phys. B61 (1973) 45.

\bibitem{Hooft2}  G. 't Hooft, Nucl. Phys. B79 (1974) 276.

\bibitem{Polyakov2}  A.M. Polyakov, JETP Lett 20 (1974) 194.

\bibitem{Duan1}  Y.S. Duan and M.L. Ge, Sci. Sinica 11 (1979) 1072.

\bibitem{Duan5}  Y.S. Duan and S.L. Zhang, Int. J. Eng. Sci. 28 (1990) 689;
29 (1991) 153; Int. J. Eng. Sci. 30 (1992) 153.

\bibitem{Duan}  Y.S. Duan, S.L. Zhang and S.S. Feng, J. Math. Phys. 35
(1994) 1; Y.S. Duan, G.H. Yang and Y. Jiang, Gen. Rel. Grav. 29 (1997) 715;
Y.S. Duan, G.H. Yang and Y. Jiang, Helv. Phys. Acta. 70 (1997) 565.

\bibitem{Duan6}  Y.S. Duan and X.H. Meng, J. Math. Phys.{\bf \ 34} 1149
(1993); Y.S. Duan and X.G. Lee, Helv. Phys. Acta. 68 (1995) 513.

\bibitem{Duan-lisheng}  Y.S. Duan and S. Li, {\em Decomposition Theory of
Spin Connection, Topological Structure of Gauss-Bonnet-Chern Topological
Current and Morse Theory}, in {\em Jingshin Theoretical Physics Symposium in
Hornor of Professor Ta-You Wu} World Scientific (1998).

\bibitem{Victor}  Victor Guillemin and Alan Pollack, in {\it Differential
Topology}, Prentice-Hall, Inc., Englewood Cliffs, New Jersey, 1974.

\bibitem{Ryder}  L.H. Ryder, J. Phys. A: Math. Gen. 13 (1980) 437.

\bibitem{Minami}  M. Minami, Prog. Theor. Phys. 62 (1979) 1128.
\end{references}
\end{document}